\documentclass[aps,prd,superscriptaddress,notitlepage,floats,floatfix,nofootinbib,amsmath,amssymb,aps]{revtex4}

\usepackage{fancyhdr}

\usepackage{bm}
\usepackage{amsfonts}
\usepackage{latexsym}
\usepackage[latin1]{inputenc}
\usepackage{graphicx}
\usepackage{amsmath}
\usepackage{palatino}
\usepackage{mathpazo}
\usepackage{booktabs}
\usepackage{dcolumn}

\def\jnl@style{\it}
\def\aaref@jnl#1{{\jnl@style#1}}

\def\aaref@jnl#1{{\jnl@style#1}}

\def\aj{\aaref@jnl{AJ}}                   
\def\apj{\aaref@jnl{ApJ}}                 
\def\apjl{\aaref@jnl{ApJ}}                
\def\apjs{\aaref@jnl{ApJS}}               
\def\apss{\aaref@jnl{Ap\&SS}}             
\def\aap{\aaref@jnl{A\&A}}                
\def\aapr{\aaref@jnl{A\&A~Rev.}}          
\def\aaps{\aaref@jnl{A\&AS}}              
\def\mnras{\aaref@jnl{Mon.~Not.~Roy.~Astron.~Soc.}}             
\def\prd{\aaref@jnl{Phys.~Rev.~D}}        
\def\prc{\aaref@jnl{Phys.~Rev.~C}}  
\def\prl{\aaref@jnl{Phys.~Rev.~Lett.}}    
\def\qjras{\aaref@jnl{QJRAS}}             
\def\skytel{\aaref@jnl{S\&T}}             
\def\ssr{\aaref@jnl{Space~Sci.~Rev.}}     
\def\zap{\aaref@jnl{ZAp}}                 
\def\nat{\aaref@jnl{Nature}}              
\def\aplett{\aaref@jnl{Astrophys.~Lett.}} 
\def\apspr{\aaref@jnl{Astrophys.~Space~Phys.~Res.}} 
\def\physrep{\aaref@jnl{Phys.~Rep.}}      
\def\physscr{\aaref@jnl{Phys.~Scr}}       
\def\commat{\aaref@jnl{Comm.~Math.~Phys.}}              
\def\science{\aaref@jnl{Science}}               
\def\cqg{\aaref@jnl{Classical Quant.~Grav.}}            
\def\jpcs{\aaref@jnl{JPCS}}                                     
\def\ijmpd{\aaref@jnl{Int.~J.~Mod.~Phys.~D}}                    
\def\grg{\aaref@jnl{Gen.~Relat.~Gravit.}}               
\def\rpp{\aaref@jnl{Rep.~Prog.~Phys.}}          
\def\npa{\aaref@jnl{Nucl.~Phys.~A}}        
\def\lrr{\aaref@jnl{Living Rev.~Rel.}}                   
\def\jcap{\aaref@jnl{J.~Cosmology Astropart.~Phys.}}    
\def\rmp{\aaref@jnl{Rev.~Mod.~Phys.}}   
\def\Universe{\aaref@jnl{Universe}}   

\usepackage{graphics}
\usepackage{graphicx}
\usepackage{color}

\allowdisplaybreaks[1]

\usepackage{soul} 

\begin{document}
\title{On Kerr black hole deformations admitting a Carter constant and an invariant criterion for the separability of the wave equation}
\author{Georgios O. Papadopoulos}
\affiliation{Department of Physics, National \& Kapodistrian University of Athens, Panepistimioupolis, GR 157-71, Athens Greece.}
\email{gopapado@phys.uoa.gr}
\author{Kostas D. Kokkotas}
\affiliation{Theoretical Astrophysics, IAAT, Eberhard Karls University of T\"ubingen, 72076, Germany}
\email{kostas.kokkotas@uni-tuebingen.de}
\date{\today}
%
\begin{abstract}

In a previous work of ours, the most general family of Kerr deformations---admitting a Carter constant--- has been presented. This time a simple,  necessary and sufficient condition in order for the aforementioned family to have a separable Klein-Gordon equations is exhibited.

\vspace{0.3cm}
\noindent
\textbf{MSC-Class} (2020): 83B05, 83C20, 83C57, 83D05, 83F05 \\
\textbf{PACS-numbers} (2010): 04.50.Kd, 04.70.Bw, 04.25.Nx, 04.30.-w, 04.80.Cc 

\end{abstract}
\maketitle
\section{Introduction}\label{Introduction}
%
During the last years an ever increasing interest on alternative variants of Black Holes (BHs) has been observed. The interest on BHs is due to not only their unique structure comprising the most extreme gravitational fields known to exist, but also their involvement in the evolution of small and large scale  structures on the universe. Indeed, the extreme gravitational fields in their vicinity make them the ideal candidates for performing tests in strong field regime. One may test not only a given quantum gravity theory, but also the uniqueness of the solutions for BHs in General Relativity (GR). If, for instance, the uniqueness of the Kerr BH is compromised then the GR, as a whole, is under question. Thus alternative variants of BHs can lead to proposals for either alternative theories for gravity (cf. references in \cite{2018CQGra..35r5014P}) or variants of the well known general relativistic metrics such as the Kerr space --see e.g., \cite{2013PhRvD..88d4002J}. These studies have been boosted by the detection of gravitational waves \cite{2016PhRvL.116f1102A,2016PhRvL.116v1101A} and the HORIZON results \cite{2019ApJ...875L...1E,2020ApJ...896....7M} which are providing promising tests of the aforementioned claims.

The great progress made during the recent years in astronomy (electromagnetic, particle and gravitational) has transformed the study of BHs from (an initially) pure academic problem into a viable and promising (both at the observational and the experimental level) research. Various theoretical models, trying to phenomenologicaly interpret the observations by LIGO/VIRGO along with the expected ones from LISA, spring out massively in the literature \cite{2019CQGra..36n3001B,2020arXiv200109793B,2020PhRvD.101j4016O}.
In principle, there are two approaches in those works; the studies are made either within the framework of a new theory for gravity (like, e.g., the $f(R)$ theory, Scalar-Tensor theories, etc) or in the context of a more pragmatic ---but still theory agnostic (like e.g., a smooth deviation from the mathematical ideal of the Kerr solution)--- approach to GR.

In both approaches, thus far, the current trend to be found in the literature is to begin with the assumption of a (mathematically convenient) phenomenological form for the metric suitable for modelling the astrophysical imprints of the source, but not (necessarily) always susceptible to some deeper analytical feature --like, for instance, the separability of a given family of equations like the wave equation etc.

In an earlier work \cite{2018CQGra..35r5014P} we introduced a simple and theory agnostic family of metrics which not only re-parameterises but also generalises many well know asymmetric metrics to be found in the literature. The novel feature of this metric is the admittance of an extra integral of motion for the corresponding geodesic equations; the well known Carter constant \cite{1968PhRv..174.1559C}.

The metric tensor field introduced in \cite{2018CQGra..35r5014P}  reads:
\begin{equation}\label{PK_metric}
		  g^{ab}=\frac{1}{A_1+B_1}\begin{pmatrix}
					 A_2 & 0 & 0 & 0\\
					 0 & B_2 & 0 & 0\\
					 0 & 0 & A_3+B_3 & A_4+B_4\\
		  0 & 0 & A_4+B_4 & A_5+B_5\end{pmatrix}
\end{equation}
where the functions $A_{i}$ depend on $r$ (i.e., ``radial functions'') and the functions $B_{i}$ depend on 
$x\equiv\cos(\theta)$ (i.e., ``angular functions''), are such that the covariant metric be asymptotically flat. This metric, by construction, allows for the existence of a second rank Killing tensor
\begin{equation}\label{PK_killing}
        K^{ab}=\frac{1}{A_1+B_1}\begin{pmatrix}
                A_2B_1 & 0 & 0 & 0\\
                0 & -B_2A_1 & 0 & 0\\
                0 & 0 & B_1A_3-A_1B_3 & B_1A_4-A_1B_4\\  
         0 & 0 & B_1A_4-A_1B_4 & B_1A_5-A_1B_5\end{pmatrix}.
\end{equation}
Actually the very existence of a Killing tensor is extremely important for it signals the formal integrability of the geodesic (or Hamilton-Jacobi) equations (cf. next Section).

The effort for construction Kerr-like metrics has been initiated more than two decades ago, but the last ten years many variations have been presented \cite{1995PhRvD..52.5707R, 2006CQGra..23.4167G,2004PhRvD..69l4022C,2011PhRvD..83l4015J, 2013PhRvD..88d4002J, 2018PhRvD..97h4044K, 2019PhRvD.100b4028S, 2020PhRvD.101h4030C}, the properties of which are compared to those of the Kerr metric. Predominantly, the interest has been related to the Hamilton-Jacobi integrability and the Klein-Gordon (KG) equation separability.

The following table shows a timely current state of affairs in the literature (only the most general results are mentioned) regarding the various alternative deviations from the Kerr metric and the integrability (via separation of the independent variables) of two important families of equations: the geodesics and the KG equation.
\begin{table}[h]
\begin{tabular}{|c||c| c| c |c|} 
 \hline
 Metric & number of                & existence of a & Geodesics & Klein-Gordon \\ 
        & free functions           & Killing tensor & separability & separability \\
 \hline\hline
 J \cite{2013PhRvD..88d4002J}      & 4 radial             & yes            & yes       & in general no \\ 
 \hline 
 KSZ \cite{2018PhRvD..97h4044K}  & 3 radial             & yes            & yes       & yes\\
 \hline
 S \cite{2019PhRvD.100b4028S}      & 3 radial             & yes            & yes       & yes\\   
 \hline 
 CY \cite{2020PhRvD.101h4030C}    & 5 radial + 1 angular & yes            & yes       & in general no \\
 \hline
 PK \cite{2018CQGra..35r5014P}    & 5 radial + 5 angular & yes            & yes       & in general no\\
 \hline
 \end{tabular}
 \caption{Comparison of the most commonly used deformations of the Kerr metric. The acronyms and references are given in the first column. The second column presents the number of assumed free deforming functions by each one of them, while the third denotes the existence of a Killing tensor while the forth comments on the separability of the geodesic equations. The fifth column, is the most relevant for the current work i.e., the separability of the KG equation.}
 \label{Tab:1}
\end{table}

At this point two remarks are in order.

First, it should be made clear that \underline{\textbf{all}} the aforementioned, 
metrics \cite{2013PhRvD..88d4002J, 2018PhRvD..97h4044K, 2019PhRvD.100b4028S, 2020PhRvD.101h4030C} are sub-cases of the PK \cite{2018CQGra..35r5014P} family of metrics; ditto for the corresponding Killing tensors.
\footnote{In \cite{2020PhRvD.101h4030C} a misunderstanding of the results published in \cite{2018CQGra..35r5014P} is apparent. More specifically, it is mentioned that according to \cite{2018CQGra..35r5014P}  the Kerr-Sen metric \cite{1992PhRvL..69.1006S} cannot be mapped to the Johannsen metric \cite{2013PhRvD..88d4002J} and that in \cite{2020PhRvD.101h4030C} it is proven to be possible. The correct statement in  \cite{2018CQGra..35r5014P} is  that the Killing tensor of the Kerr-Sen metric, which is of Petrov type I, is not induced by a Yano tensor --and thus, it is somehow more general than the rest of the examples, which are of the Petrov type D. Never the less the Johannsen metric is ---in priciple--- of Petrov type I as well and contains much freedom so that the Kerr-Sen metric can be considered as a member of the this family.}

Second, barring the PK \cite{2018CQGra..35r5014P} metric, all the others ---to some extent--- are related to some theory (e.g., the J metric \cite{2013PhRvD..88d4002J} is a pragmatic extension of GR, etc). Indeed, a deformed Kerr metric can be obtained by two ways: 
\begin{itemize}
\item either one implements the equivalent of a kinematical algorithm --something which does map solutions to 
the Einstein Field Equations (EFEs) to other solutions (to the EFEs); like the S metric 
\cite{2019PhRvD.100b4028S} where the author starts with a non rotating solution to the EFEs, then exploits the 
Newman-Janis (complex) algorithm and ends up with another solution (to the same theory, i.e., GR)
\item or one considers a completely different metric, deviating smoothly from the Kerr solution, which has to 
satisfy some field equations (e.g., the GR plus some perturbation of the energy-momentum tensor see, e.g., 
\cite{2019PhRvD..99l4026S}, \cite{2020PhRvD.101h4030C} and references therein)
\end{itemize}
On the contrary, the PK family of metrics \cite{2018CQGra..35r5014P} was constructed as a geometric object endowed with the maximum number of free functions along with some specific geometric features (like the existence of a Killing tensor, something very important for this leads to integrals of motion for the geodesic equations) without any reference to any theory. Of course there is a price for this generality; e.g., not all members of the family are ---in principle--- physically accepted; supplementary constraints like e.g., the energy conditions, have to be implemented when one considers any member of this family.

In the present work a further step is made and a new family of metrics, \emph{we will call it PK-I family},  is presented. The new family is a \emph{subfamily} of the PK one \cite{2018CQGra..35r5014P}. Indeed, the new family (PK-I) consists of the members of the previous one (PK) obeying  a necessary and sufficient condition (upon the metric functions) such that the Klein-Gordon equation be separable.

It should be noted that classical references on the separability of the KG and its generalisation on the Kerr background are to be found in \cite{1968PhRv..174.1559C, 1972PhRvD...5.1913B, 1972PhRvL..29.1114T}. 
Also, in \cite{2019PhRvD.100j4054C} a very interesting  comparable analysis of the KG separability was presented  for various alternative phenomenological families of metrics. 

In the following section a formal description of the PK-I (sub)family of metric spaces will be presented, based solely on an implicit yet invariant criterion. An immediate yet fundamental application of this will be exhibited.

\section{PK-I metrics: formal existence and an application}

In general, separability structures are closely related to the existence of the so called (possibly hidden) symmetries; a kind of physical degeneracy (e.g., cf. N\"{o}ther's theorem, and \cite{1975ctf..book.....L}). 
When it comes to gravitation the studies focus mainly on the geodesic equations. Never the less, there are other equations of physical interest as well. Such an example is the KG equation, which is also a very
useful tool in observational physics. So it would be interesting if one could somehow enhance the previous considerations ---on symmetries and integrability--- to the case of KG equation.

Let assume the metric $g^{ab}$ as it is given in \eqref{PK_metric}. Then this metric has the fundamental property of admitting a Carter constant. Indeed, let construct the scalar functional
\begin{equation}
	I=K_{ab}\dot{x}^{a}\dot{x}^{b}		  
\end{equation}
with the Killing tensor obeying, by definition, the condition
\begin{equation}\label{Killing_Tensor_Property}
					     \nabla_{(a}K_{bc)}=0
 \end{equation}
 then $I$ is a constant of motion of the geodesics equations
\begin{equation}\label{geodesics}                                \ddot{x}^{c}+\Gamma^{c}_{\phantom{1}ab}\dot{x}^{a}\dot{x}^{b}=0
\end{equation}
 since the combination of \eqref{Killing_Tensor_Property} and \eqref{geodesics} leads to
\begin{equation}
\dot{I}={\nabla}_{c}K_{ab}\dot{x}^{a}\dot{x}^{b}\dot{x}^{c}=0
\end{equation}
i.e., the last quantity vanishes by virtue of the geodesics and the Killing tensor definition. 

Now the focus is on the Klein-Gordon equation for a scalar field $\Psi$ on the background of the aforementioned metric: 
\begin{equation}
	\nabla^{a}\nabla_{a}\Psi=0 \, .
 \end{equation}
A tedious yet straightforward calculation shows that this equation is susceptible to separation of variables if under the Ansatz
 \begin{equation}
 \Psi(t,r,x,\phi)=e^{i(m\phi-\omega t)}{\cal X}(r) \, {\cal Y}(x)
 \end{equation}
it holds that 
\begin{equation}    
\nabla^{a}\nabla_{a}\Psi=0 \Rightarrow 
\text{Radial Part}+\text{Angular Part}=0
\end{equation}
where (modulo overall factors) 
\begin{eqnarray}
\text{Radial Part}&:& 
                   A_2{\partial_{rr}{\cal X}}+\Big(\partial_r A_2+A_2\partial_r\Omega\Big)
								{\partial_r {\cal X}}-\left(m^2A_3-2m\omega A_4+\omega^2 A_5+\lambda\right){\cal X} = 0
								\label{eq:Radial}
								\\
\text{Angular Part}&:& 
B_2{\partial_{xx}{\cal Y}}+\Big(\partial_x B_2+B_2\partial_x\Omega\Big)
                        \partial_x {\cal Y}-\left(m^2B_3-2m\omega B_4+\omega^2 B_5-\lambda\right) {\cal Y} = 0
                        \label{eq:Angular}
 \end{eqnarray}
with $\lambda$ denoting the separation constant, $g$ the determinant of the metric tensor\footnote{The exact form of $g$  is quite extended to be written here.}, while the quantity 
$\Omega$ is identified as
\begin{equation}
\Omega=\ln\Big(\frac{\sqrt{-g}}{(A_1(r)+B_1(x))}\Big) \, .
\end{equation}

Now, it is obvious that the Radial and the Angular parts separate \underline{\textbf{if and only if}}:
\begin{equation}
  \Omega=\mathcal{F}_1(r)+\mathcal{F}_2(x)  
\end{equation}
where $\mathcal{F}_1, \mathcal{F}_2$ are supposed to be well behaved functions of their designated arguments.
Just for reference, in the case of the Kerr metric itself it is $\Omega_{\text{Kerr}}=0$ while in the case of the Kerr-Sen metric \cite{1992PhRvL..69.1006S} it is $\Omega_{\text{Kerr-Sen}}=1$.

An interesting application of the previous consideration would be the counting of the maximum number of free radial functions, when the angular functions are taken to be the Kerrian ones:
\begin{subequations}
\begin{align}
      B_1 &= a^2 x^2\\
      B_2 &= 1-x^2\\
      B_3 &= \frac{1}{1-x^2}\\
      B_4 &= a\\
      B_5 &= a^2(1-x^2) \, .
\end{align}
\end{subequations}
The necessary and sufficient condition, regarding $\Omega$ becomes a polynomial constraint, in the angular variable $x$. The vanishing of the coefficients results in \emph{three free radial functions}: $A_2, A_3$ and any of rest three (i.e., any one of $A_1, A_4, A_5$), as the desired maximum number. Actually,  the three \emph{free} functions are completely at one's disposal, while the rest two are related to those three. This result is in full agreement with the existing relevant attempts to be found in the recent literature --cf. \cite{2018PhRvD..97h4044K} and \cite{2019PhRvD.100b4028S}.
Indeed, for reference \cite{2018PhRvD..97h4044K} it is:
\begin{subequations}
\begin{align}
A_1 &= r^2 R_\Sigma(r)\\
A_2 &=\frac{a^2-r R_{M}(r)+r^2 R_{\Sigma}(r)}{R_B (r)^2}\\
A_3 &=-\frac{a^2}{a^2-r R_{M}(r)+r^2 R_{\Sigma}(r)}\\
A_4 &=-\frac{a^3+a r^2 R_{\Sigma}(r)}{a^2-r R_{M}(r)+r^2 R_{\Sigma}(r)}\\
A_5 &=-\frac{(a^2+r^2 R_{\Sigma}(r))^2}{a^2-r R_{M}(r)+r^2 R_{\Sigma}(r)}\\
\Omega &=\ln\Big(R_{B}(r)\Big)
\end{align}
\end{subequations}
while for reference \cite{2019PhRvD.100b4028S} it is:
\begin{subequations}
\begin{align}
A_2 &= \Delta(r)\\
A_3 &= -\frac{a^2}{\Delta(r)}\\
A_4 &= -\frac{a X(r)}{\Delta(r)}\\
A_5 &= -\frac{X(r)^2}{\Delta(r)}\\
\Omega &=\ln\Big(\frac{a^2x^2+A_1(r)}{-a^2(1-x^2)+X(r)}\Big)
\end{align}
\end{subequations}
and this time, $A_1(r)$ is a proper function such that the corresponding $\Omega$ be separable as a function.

\section{Discussion}

In this short work, we have reported on those (sub)families of space times which constitute (smooth) Kerr deformations endowed with the properties of
\begin{itemize}
    \item admitting a Carter constant --something signaling the separability and the complete
          integration of the geodesic equations
    \item allowing for the separability of the KG equation.
\end{itemize}
This (sub)family is given rather indirectly, through an invariant criterion, never the less the result itself is important per se for, two simple reasons:
\begin{enumerate}
    \item it confirms that for the Kerrian choice of radial functions, the maximum freedom
          of the radial functions is three --something which not only confirms other, less general attempts in the recent literature but it inherits to them a solid theoretical foundation.
    \item it provides, the most general subfamily of those space-time smooth Kerr deformations
          which on one hand admit a Carter constant and on the other hand the KG equation
          is separable with the freedom of (at most) ten free functions obying a single constraint.
\end{enumerate}

\acknowledgements{The authors are grateful to  K.~Destounis, S.~Nampalliwar and A.~G.~Suvorov for useful discussions. This work was supported by the DAAD program  ``Hochschulpartnerschaften mit Griechenland 2016'' (Projekt 57340132). Networking support by the COST Action  CA16104  is also gratefully acknowledged.
}

\bibliography{references}

\begin{thebibliography}{23}
\expandafter\ifx\csname natexlab\endcsname\relax\def\natexlab#1{#1}\fi
\expandafter\ifx\csname bibnamefont\endcsname\relax
  \def\bibnamefont#1{#1}\fi
\expandafter\ifx\csname bibfnamefont\endcsname\relax
  \def\bibfnamefont#1{#1}\fi
\expandafter\ifx\csname citenamefont\endcsname\relax
  \def\citenamefont#1{#1}\fi
\expandafter\ifx\csname url\endcsname\relax
  \def\url#1{\texttt{#1}}\fi
\expandafter\ifx\csname urlprefix\endcsname\relax\def\urlprefix{URL }\fi
\providecommand{\bibinfo}[2]{#2}
\providecommand{\eprint}[2][]{\url{#2}}

\bibitem[{\citenamefont{{Papadopoulos} and
  {Kokkotas}}(2018)}]{2018CQGra..35r5014P}
\bibinfo{author}{\bibfnamefont{G.~O.} \bibnamefont{{Papadopoulos}}}
  \bibnamefont{and} \bibinfo{author}{\bibfnamefont{K.~D.}
  \bibnamefont{{Kokkotas}}}, \bibinfo{journal}{\cqg}
  \textbf{\bibinfo{volume}{35}}, \bibinfo{eid}{185014} (\bibinfo{year}{2018}),
  \eprint{1807.08594}.

\bibitem[{\citenamefont{{Johannsen}}(2013)}]{2013PhRvD..88d4002J}
\bibinfo{author}{\bibfnamefont{T.}~\bibnamefont{{Johannsen}}},
  \bibinfo{journal}{\prd} \textbf{\bibinfo{volume}{88}}, \bibinfo{eid}{044002}
  (\bibinfo{year}{2013}), \eprint{1501.02809}.

\bibitem[{\citenamefont{{Abbott}
  et~al.}(2016{\natexlab{a}})\citenamefont{{Abbott}, {Abbott}, {Abbott},
  {Abernathy}, {Acernese}, {Ackley}, {Adams}, {Adams}, {Addesso}, {Adhikari}
  et~al.}}]{2016PhRvL.116f1102A}
\bibinfo{author}{\bibfnamefont{B.~P.} \bibnamefont{{Abbott}}},
  \bibinfo{author}{\bibfnamefont{R.}~\bibnamefont{{Abbott}}},
  \bibinfo{author}{\bibfnamefont{T.~D.} \bibnamefont{{Abbott}}},
  \bibinfo{author}{\bibfnamefont{M.~R.} \bibnamefont{{Abernathy}}},
  \bibinfo{author}{\bibfnamefont{F.}~\bibnamefont{{Acernese}}},
  \bibinfo{author}{\bibfnamefont{K.}~\bibnamefont{{Ackley}}},
  \bibinfo{author}{\bibfnamefont{C.}~\bibnamefont{{Adams}}},
  \bibinfo{author}{\bibfnamefont{T.}~\bibnamefont{{Adams}}},
  \bibinfo{author}{\bibfnamefont{P.}~\bibnamefont{{Addesso}}},
  \bibinfo{author}{\bibfnamefont{R.~X.} \bibnamefont{{Adhikari}}},
  \bibnamefont{et~al.}, \bibinfo{journal}{\prl} \textbf{\bibinfo{volume}{116}},
  \bibinfo{eid}{061102} (\bibinfo{year}{2016}{\natexlab{a}}),
  \eprint{1602.03837}.

\bibitem[{\citenamefont{{Abbott}
  et~al.}(2016{\natexlab{b}})\citenamefont{{Abbott}, {Abbott}, {Abbott},
  {Abernathy}, {Acernese}, {Ackley}, {Adams}, {Adams}, {Addesso}, {Adhikari}
  et~al.}}]{2016PhRvL.116v1101A}
\bibinfo{author}{\bibfnamefont{B.~P.} \bibnamefont{{Abbott}}},
  \bibinfo{author}{\bibfnamefont{R.}~\bibnamefont{{Abbott}}},
  \bibinfo{author}{\bibfnamefont{T.~D.} \bibnamefont{{Abbott}}},
  \bibinfo{author}{\bibfnamefont{M.~R.} \bibnamefont{{Abernathy}}},
  \bibinfo{author}{\bibfnamefont{F.}~\bibnamefont{{Acernese}}},
  \bibinfo{author}{\bibfnamefont{K.}~\bibnamefont{{Ackley}}},
  \bibinfo{author}{\bibfnamefont{C.}~\bibnamefont{{Adams}}},
  \bibinfo{author}{\bibfnamefont{T.}~\bibnamefont{{Adams}}},
  \bibinfo{author}{\bibfnamefont{P.}~\bibnamefont{{Addesso}}},
  \bibinfo{author}{\bibfnamefont{R.~X.} \bibnamefont{{Adhikari}}},
  \bibnamefont{et~al.}, \bibinfo{journal}{\prl} \textbf{\bibinfo{volume}{116}},
  \bibinfo{eid}{221101} (\bibinfo{year}{2016}{\natexlab{b}}),
  \eprint{1602.03841}.

\bibitem[{\citenamefont{{Event Horizon Telescope Collaboration}
  et~al.}(2019)\citenamefont{{Event Horizon Telescope Collaboration},
  {Akiyama}, {Alberdi}, {Alef}, {Asada}, {Azulay}, {Baczko}, {Ball},
  {Balokovi{\'c}}, {Barrett} et~al.}}]{2019ApJ...875L...1E}
\bibinfo{author}{\bibnamefont{{Event Horizon Telescope Collaboration}}},
  \bibinfo{author}{\bibfnamefont{K.}~\bibnamefont{{Akiyama}}},
  \bibinfo{author}{\bibfnamefont{A.}~\bibnamefont{{Alberdi}}},
  \bibinfo{author}{\bibfnamefont{W.}~\bibnamefont{{Alef}}},
  \bibinfo{author}{\bibfnamefont{K.}~\bibnamefont{{Asada}}},
  \bibinfo{author}{\bibfnamefont{R.}~\bibnamefont{{Azulay}}},
  \bibinfo{author}{\bibfnamefont{A.-K.} \bibnamefont{{Baczko}}},
  \bibinfo{author}{\bibfnamefont{D.}~\bibnamefont{{Ball}}},
  \bibinfo{author}{\bibfnamefont{M.}~\bibnamefont{{Balokovi{\'c}}}},
  \bibinfo{author}{\bibfnamefont{J.}~\bibnamefont{{Barrett}}},
  \bibnamefont{et~al.}, \bibinfo{journal}{\apjl}
  \textbf{\bibinfo{volume}{875}}, \bibinfo{eid}{L1} (\bibinfo{year}{2019}),
  \eprint{1906.11238}.

\bibitem[{\citenamefont{{Medeiros} et~al.}(2020)\citenamefont{{Medeiros},
  {Psaltis}, and {{\"O}zel}}}]{2020ApJ...896....7M}
\bibinfo{author}{\bibfnamefont{L.}~\bibnamefont{{Medeiros}}},
  \bibinfo{author}{\bibfnamefont{D.}~\bibnamefont{{Psaltis}}},
  \bibnamefont{and}
  \bibinfo{author}{\bibfnamefont{F.}~\bibnamefont{{{\"O}zel}}},
  \bibinfo{journal}{\apj} \textbf{\bibinfo{volume}{896}}, \bibinfo{eid}{7}
  (\bibinfo{year}{2020}), \eprint{1907.12575}.

\bibitem[{\citenamefont{{Barack} et~al.}(2019)\citenamefont{{Barack},
  {Cardoso}, {Nissanke}, {Sotiriou}, {Askar}, {Belczynski}, {Bertone}, {Bon},
  {Blas}, {Brito} et~al.}}]{2019CQGra..36n3001B}
\bibinfo{author}{\bibfnamefont{L.}~\bibnamefont{{Barack}}},
  \bibinfo{author}{\bibfnamefont{V.}~\bibnamefont{{Cardoso}}},
  \bibinfo{author}{\bibfnamefont{S.}~\bibnamefont{{Nissanke}}},
  \bibinfo{author}{\bibfnamefont{T.~P.} \bibnamefont{{Sotiriou}}},
  \bibinfo{author}{\bibfnamefont{A.}~\bibnamefont{{Askar}}},
  \bibinfo{author}{\bibfnamefont{C.}~\bibnamefont{{Belczynski}}},
  \bibinfo{author}{\bibfnamefont{G.}~\bibnamefont{{Bertone}}},
  \bibinfo{author}{\bibfnamefont{E.}~\bibnamefont{{Bon}}},
  \bibinfo{author}{\bibfnamefont{D.}~\bibnamefont{{Blas}}},
  \bibinfo{author}{\bibfnamefont{R.}~\bibnamefont{{Brito}}},
  \bibnamefont{et~al.}, \bibinfo{journal}{Classical and Quantum Gravity}
  \textbf{\bibinfo{volume}{36}}, \bibinfo{eid}{143001} (\bibinfo{year}{2019}),
  \eprint{1806.05195}.

\bibitem[{\citenamefont{{Barausse} et~al.}(2020)\citenamefont{{Barausse},
  {Berti}, {Hertog}, {Hughes}, {Jetzer}, {Pani}, {Sotiriou}, {Tamanini},
  {Witek}, {Yagi} et~al.}}]{2020arXiv200109793B}
\bibinfo{author}{\bibfnamefont{E.}~\bibnamefont{{Barausse}}},
  \bibinfo{author}{\bibfnamefont{E.}~\bibnamefont{{Berti}}},
  \bibinfo{author}{\bibfnamefont{T.}~\bibnamefont{{Hertog}}},
  \bibinfo{author}{\bibfnamefont{S.~A.} \bibnamefont{{Hughes}}},
  \bibinfo{author}{\bibfnamefont{P.}~\bibnamefont{{Jetzer}}},
  \bibinfo{author}{\bibfnamefont{P.}~\bibnamefont{{Pani}}},
  \bibinfo{author}{\bibfnamefont{T.~P.} \bibnamefont{{Sotiriou}}},
  \bibinfo{author}{\bibfnamefont{N.}~\bibnamefont{{Tamanini}}},
  \bibinfo{author}{\bibfnamefont{H.}~\bibnamefont{{Witek}}},
  \bibinfo{author}{\bibfnamefont{K.}~\bibnamefont{{Yagi}}},
  \bibnamefont{et~al.}, \bibinfo{journal}{arXiv e-prints}
  \bibinfo{eid}{arXiv:2001.09793} (\bibinfo{year}{2020}), \eprint{2001.09793}.

\bibitem[{\citenamefont{{Okounkova} et~al.}(2020)\citenamefont{{Okounkova},
  {Stein}, {Moxon}, {Scheel}, and {Teukolsky}}}]{2020PhRvD.101j4016O}
\bibinfo{author}{\bibfnamefont{M.}~\bibnamefont{{Okounkova}}},
  \bibinfo{author}{\bibfnamefont{L.~C.} \bibnamefont{{Stein}}},
  \bibinfo{author}{\bibfnamefont{J.}~\bibnamefont{{Moxon}}},
  \bibinfo{author}{\bibfnamefont{M.~A.} \bibnamefont{{Scheel}}},
  \bibnamefont{and} \bibinfo{author}{\bibfnamefont{S.~A.}
  \bibnamefont{{Teukolsky}}}, \bibinfo{journal}{\prd}
  \textbf{\bibinfo{volume}{101}}, \bibinfo{eid}{104016} (\bibinfo{year}{2020}),
  \eprint{1911.02588}.

\bibitem[{\citenamefont{{Carter}}(1968)}]{1968PhRv..174.1559C}
\bibinfo{author}{\bibfnamefont{B.}~\bibnamefont{{Carter}}},
  \bibinfo{journal}{Physical Review} \textbf{\bibinfo{volume}{174}},
  \bibinfo{pages}{1559} (\bibinfo{year}{1968}).

\bibitem[{\citenamefont{{Ryan}}(1995)}]{1995PhRvD..52.5707R}
\bibinfo{author}{\bibfnamefont{F.~D.} \bibnamefont{{Ryan}}},
  \bibinfo{journal}{\prd} \textbf{\bibinfo{volume}{52}}, \bibinfo{pages}{5707}
  (\bibinfo{year}{1995}).

\bibitem[{\citenamefont{{Glampedakis} and {Babak}}(2006)}]{2006CQGra..23.4167G}
\bibinfo{author}{\bibfnamefont{K.}~\bibnamefont{{Glampedakis}}}
  \bibnamefont{and} \bibinfo{author}{\bibfnamefont{S.}~\bibnamefont{{Babak}}},
  \bibinfo{journal}{\cqg} \textbf{\bibinfo{volume}{23}}, \bibinfo{pages}{4167}
  (\bibinfo{year}{2006}), \eprint{gr-qc/0510057}.

\bibitem[{\citenamefont{{Collins} and {Hughes}}(2004)}]{2004PhRvD..69l4022C}
\bibinfo{author}{\bibfnamefont{N.~A.} \bibnamefont{{Collins}}}
  \bibnamefont{and} \bibinfo{author}{\bibfnamefont{S.~A.}
  \bibnamefont{{Hughes}}}, \bibinfo{journal}{\prd}
  \textbf{\bibinfo{volume}{69}}, \bibinfo{eid}{124022} (\bibinfo{year}{2004}),
  \eprint{gr-qc/0402063}.

\bibitem[{\citenamefont{{Johannsen} and {Psaltis}}(2011)}]{2011PhRvD..83l4015J}
\bibinfo{author}{\bibfnamefont{T.}~\bibnamefont{{Johannsen}}} \bibnamefont{and}
  \bibinfo{author}{\bibfnamefont{D.}~\bibnamefont{{Psaltis}}},
  \bibinfo{journal}{\prd} \textbf{\bibinfo{volume}{83}}, \bibinfo{eid}{124015}
  (\bibinfo{year}{2011}), \eprint{1105.3191}.

\bibitem[{\citenamefont{{Konoplya} et~al.}(2018)\citenamefont{{Konoplya},
  {Stuchl{\'\i}k}, and {Zhidenko}}}]{2018PhRvD..97h4044K}
\bibinfo{author}{\bibfnamefont{R.~A.} \bibnamefont{{Konoplya}}},
  \bibinfo{author}{\bibfnamefont{Z.}~\bibnamefont{{Stuchl{\'\i}k}}},
  \bibnamefont{and}
  \bibinfo{author}{\bibfnamefont{A.}~\bibnamefont{{Zhidenko}}},
  \bibinfo{journal}{\prd} \textbf{\bibinfo{volume}{97}}, \bibinfo{eid}{084044}
  (\bibinfo{year}{2018}), \eprint{1801.07195}.

\bibitem[{\citenamefont{{Shaikh}}(2019)}]{2019PhRvD.100b4028S}
\bibinfo{author}{\bibfnamefont{R.}~\bibnamefont{{Shaikh}}},
  \bibinfo{journal}{\prd} \textbf{\bibinfo{volume}{100}}, \bibinfo{eid}{024028}
  (\bibinfo{year}{2019}), \eprint{1904.08322}.

\bibitem[{\citenamefont{{Carson} and {Yagi}}(2020)}]{2020PhRvD.101h4030C}
\bibinfo{author}{\bibfnamefont{Z.}~\bibnamefont{{Carson}}} \bibnamefont{and}
  \bibinfo{author}{\bibfnamefont{K.}~\bibnamefont{{Yagi}}},
  \bibinfo{journal}{\prd} \textbf{\bibinfo{volume}{101}}, \bibinfo{eid}{084030}
  (\bibinfo{year}{2020}), \eprint{2002.01028}.

\bibitem[{\citenamefont{{Sen}}(1992)}]{1992PhRvL..69.1006S}
\bibinfo{author}{\bibfnamefont{A.}~\bibnamefont{{Sen}}},
  \bibinfo{journal}{\prl} \textbf{\bibinfo{volume}{69}}, \bibinfo{pages}{1006}
  (\bibinfo{year}{1992}), \eprint{hep-th/9204046}.

\bibitem[{\citenamefont{{Suvorov}}(2019)}]{2019PhRvD..99l4026S}
\bibinfo{author}{\bibfnamefont{A.~G.} \bibnamefont{{Suvorov}}},
  \bibinfo{journal}{\prd} \textbf{\bibinfo{volume}{99}}, \bibinfo{eid}{124026}
  (\bibinfo{year}{2019}), \eprint{1905.02021}.

\bibitem[{\citenamefont{{Brill} et~al.}(1972)\citenamefont{{Brill},
  {Chrzanowski}, {Pereira}, {Fackerell}, and {Ipser}}}]{1972PhRvD...5.1913B}
\bibinfo{author}{\bibfnamefont{D.~R.} \bibnamefont{{Brill}}},
  \bibinfo{author}{\bibfnamefont{P.~L.} \bibnamefont{{Chrzanowski}}},
  \bibinfo{author}{\bibfnamefont{C.~M.} \bibnamefont{{Pereira}}},
  \bibinfo{author}{\bibfnamefont{E.~D.} \bibnamefont{{Fackerell}}},
  \bibnamefont{and} \bibinfo{author}{\bibfnamefont{J.~R.}
  \bibnamefont{{Ipser}}}, \bibinfo{journal}{\prd} \textbf{\bibinfo{volume}{5}},
  \bibinfo{pages}{1913} (\bibinfo{year}{1972}).

\bibitem[{\citenamefont{{Teukolsky}}(1972)}]{1972PhRvL..29.1114T}
\bibinfo{author}{\bibfnamefont{S.~A.} \bibnamefont{{Teukolsky}}},
  \bibinfo{journal}{\prl} \textbf{\bibinfo{volume}{29}}, \bibinfo{pages}{1114}
  (\bibinfo{year}{1972}).

\bibitem[{\citenamefont{{Chen} and {Chen}}(2019)}]{2019PhRvD.100j4054C}
\bibinfo{author}{\bibfnamefont{C.-Y.} \bibnamefont{{Chen}}} \bibnamefont{and}
  \bibinfo{author}{\bibfnamefont{P.}~\bibnamefont{{Chen}}},
  \bibinfo{journal}{\prd} \textbf{\bibinfo{volume}{100}}, \bibinfo{eid}{104054}
  (\bibinfo{year}{2019}), \eprint{1909.06968}.

\bibitem[{\citenamefont{{Landau} and {Lifshitz}}(1975)}]{1975ctf..book.....L}
\bibinfo{author}{\bibfnamefont{L.~D.} \bibnamefont{{Landau}}} \bibnamefont{and}
  \bibinfo{author}{\bibfnamefont{E.~M.} \bibnamefont{{Lifshitz}}},
  \emph{\bibinfo{title}{{The classical theory of fields}}}
  (\bibinfo{publisher}{Pergamon Press}, \bibinfo{year}{1975}).

\end{thebibliography}

\end{document}